%% file: paper.tex
\documentclass[sigconf]{acmart}
\usepackage{multirow, subcaption}
\usepackage{tabularx} 
\usepackage[table]{xcolor}
\definecolor{lightgray}{gray}{0.95}

\newcolumntype{Y}{>{\raggedright\arraybackslash}X}

\AtBeginDocument{%
  }

\begin{document}

\title{Social Simulation for Mental Health: Evaluating Contextual Environments in Augmented Reality for Training Self-Care Skills}

\author{Anna Fang}
\affiliation{%
  \institution{Carnegie Mellon University}
  \city{Pittsburgh}
  \state{Pennsylvania}
  \country{USA}
}

\author{Jiayang Shi}
\affiliation{%
  \institution{Carnegie Mellon University}
  \city{Pittsburgh}
  \state{Pennsylvania}
  \country{USA}
}

\author{Hriday Chhabria}
\affiliation{%
  \institution{University of Michigan}
  \city{Ann Arbor}
  \state{Michigan}
  \country{USA}
}

\author{Haiyi Zhu}
\affiliation{%
  \institution{Carnegie Mellon University}
  \city{Pittsburgh}
  \state{Pennsylvania}
  \country{USA}
}

\renewcommand{\shortauthors}{Fang et al.}

\begin{abstract}
Stress and anxiety are common, and there is growing interest in using VR/AR for training coping skills. However, these skills are meant to be applied during real-world distress, while existing interventions largely focus on calm or abstract settings. Little is known about how simulating realistic environments affects stress responses and skill transfer. We developed an AR social simulation intervention for practicing self-care skills, and contribute a 14-day, 43-participant experiment to examine its effects versus a system without social simulation. Results showed that participants did not differ in frequency of intervention use nor self-reported stress, but those in the social simulation condition exhibited slower heart-rate rise during a speaking task, steeper rate of heart-rate decline when self-soothing, and more frequent application of skills in the real-world dependent on if opportunities arose. Our work contributes empirical understanding of realistic environments for mental health training, and implications for designing embodied interventions for self-care. \end{abstract}



\begin{teaserfigure}
  \includegraphics[width=\linewidth]{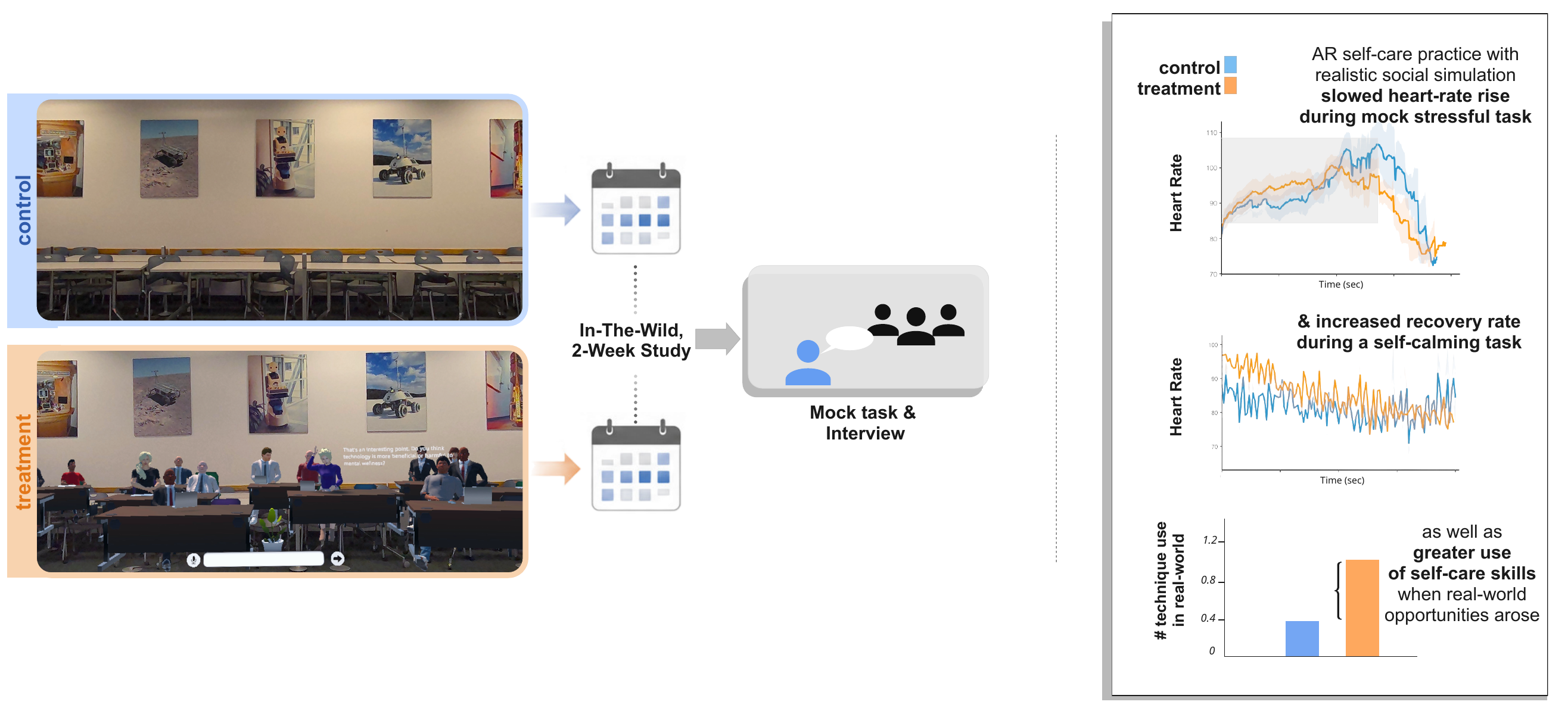}
  \caption{We present findings from a randomized \textbf{14-day, 43-participant experiment} comparing training of self-soothing skills using an integrated AR social simulation versus a control. We compare how a more realistic stressor environment for practice influences outcomes of usage, effectiveness, and skill transfer, applied to the context of public speaking.}
  \Description{A study flow showing control and treatment versions from our developed AR system, which is then used for an in-the-wild, 2-week study. We evaluate how participants perform when self-soothing through a mock speaking task and derive qualitative insight through a semi-structured interview. Shown on the right-hand side are quantitative results reveals that participants who trained in an AR self-care practice showed slower modeled heart-rate rise during the stress task, increasing recovery rate during self-soothing, and greater use of skills in their real-world lives dependent on if opportunities arose.}
  \label{fig:teaser}
\end{teaserfigure}

\maketitle
\input{Sections/01-introduction}
\input{Sections/02-relatedwork}
\input{Sections/03-hypotheses}
\input{Sections/04-methods}

\input{Sections/05-results}

\input{Sections/06-discussion}
\input{Sections/07-conclusion}
\bibliographystyle{ACM-Reference-Format}

\bibliography{biblio}

\appendix
\section{Prompting for avatar interaction with gpt4o}

\textit{I want to simulate being questioned by an audience of people about my speech. There should be uncomfortable or awkward moments in the simulation, as there often are at public speaking events, such as audience members sometimes being skeptical or dismissive. The topic for my speech is [TOPIC].}

\textit{I should be able to converse with 15 people in the simulation total. There are 15 IDs, ranging from 0 to 14, and one person can say dialogue more than once. Do not return two personas' dialogues because only one person can ask a question at a time.}

\textit{When I enter ‘BEGIN’: Start the simulation. I will play myself, and you will play the audience members, who are based on the personas you came up with. Audience members should question me about my speech topic. It is very important that you never have two people ask me questions or speak at the same time.}

\textit{Simulate the other people's dialogue using the following JSON format:
{""id"": ""Person ID Here"",
""Content"": ""Dialogue Here""
}}

\textit{Additional Notes:
- Respond ONLY with the JSON. Don't include any header text that formats the json, just provide me the list as if you were a REST API.
- If the user asks 'Could I take a moment?', always output a JSON of a persona saying something along the lines of 'Sure' or 'Yes, just respond to the question when you are ready'.
- Please be sure to use double quotes for json objects "".
- For the dialogue, please just return a single json object, never multiple.
- IDs start at 0. (0-Index)
- Always return in the above JSON format. Even if the user says something unexpected, just say something in the above format.";}

\section{Full Daily Report Survey for Participants}

\begin{enumerate}
    \item (Before you start the application)
Have you used any self-soothing skills in your life outside of the AR headset in the past 24-hours?
\item (Before you start the application)
What is your stress level at the current moment?
\item (Before you start the application)
What is your anticipated stress level, if you had to conduct public speaking in real life?
\item (After you end the application)
What is your overall stress level at the current moment?
\item (After you end the application)
What is your anticipated stress level, if you had to conduct public speaking in real life?
\end{enumerate}
 Questions 2 and 4 regarding stress at the current moment have multiple choice options of Likert scale 1 (Not at all stressed) to 5 (Extremely Stressed), while Questions 3 and 5 have the SUDS scale options of:
 
"\textit{\textbf{0}:   No distress at all. Completely relaxed.; 
\textbf{1}:   No acute distress. Alert, focused, calm.;
\textbf{2}:   Minimal distress. Not noticeable unless I think deeply about it.;
\textbf{3}:   Mild distress. Slightly worried, but can still focus without issue.;
\textbf{4}:   Mild to moderate distress. Can still function normally, but unpleasant feeling is harder to ignore.;
\textbf{5}:   Moderate distress. Experiencing distress -- but unpleasant feelings are manageable with some effort.;
\textbf{6}:   Moderate to strong distress. Feeling bad enough that I feel like something should be done.;
\textbf{7}:   Moderately high distress. Feeling tense and losing focus. I can maintain control but with some difficulty.;
\textbf{8}:   High distress. High level of uncomfortable emotions, which cannot be tolerated for too long.;
\textbf{9}:   High to extreme distress. Losing control of emotions, feeling desperate, at risk of poor choices.;
\textbf{10}:   Extreme distress. Out of control, cannot function. Unable to think clearly, only react.}"

\section{Heart-Rate Variability Results}



\begin{table}[ht]
\centering
\begin{tabular}{lcc}
\toprule
\textbf{Variable} & \shortstack{\textbf{(a) SDNN (30s)} \\ $\beta$ (SE)} & \shortstack{\textbf{(b) RMSSD (30s)} \\ $\beta$ (SE)} \\
\midrule
(Intercept)            & 52.99*** (14.34) & 15.27** (7.28) \\
Treatment              & -2.56 (2.73)     & 0.28 (1.39) \\
Time (sec)             & -0.0051 (0.0033) & 0.0132*** (0.0017) \\
Resting BPM            & -0.367 (0.182) & -0.120 (0.092) \\
Treatment $\times$ Time & -0.00002 (0.0043) & -0.0127 (0.0023) \\
\midrule
\textbf{Random effects} & \multicolumn{2}{c}{Random intercept for participant ID} \\
\bottomrule
\end{tabular}
\caption{Linear mixed-effects models predicting HRV measures during the speaking task (30-second windows). Model (a) uses SDNN, model (b) uses RMSSD. Significance: *** $p<0.001$, ** $p<0.01$, * $p<0.05$}
\label{table:hrv_models}
\end{table}

\end{document}

%% file: Sections/01-introduction.tex
\section{Introduction}
Over the past several decades, human-computer interaction (HCI) has become increasingly interested in designing immersive and interactive tools to support mental health and well-being. For example, virtual and augmented reality (VR/AR) applications provide embodied environments that can help people experience calming or therapeutic settings \cite{riches2024virtual, freeman2017virtual}. This trend in HCI is informed by research in psychology, education, and health, which has all demonstrated that embodied and immersive interaction can foster more engaging and adaptive skill development \cite{xie2008comparing, marshall2003conceptualising, 10.1145/1347390.1347433}. Building on these foundations, HCI research for mental health has used a range of interactive stimuli in order to enhance presence, support behavior change, and improve users’ well-being.

However, despite the numerous benefits of immersive technology for mental health such as greater engagement, better learning outcomes, and more enjoyment from users \cite{wang2022reducing, freeman2017virtual}, these interventions largely remain focused on short-term relief and aiding users by presenting calm \cite{chandrasiri2020virtual, miner2024stairway}, otherworldly spaces \cite{miller2023awedyssey}, abstract \cite{ng2023virtual}, and gamified \cite{wang2023breathero} environments. These settings are effective for immediate stress reduction, but are disconnected from the everyday contexts in which mental distress arises -- e.g., at work, in our personal relationships, at social events. As a result, users may struggle to actually apply self-care or soothing strategies when confronted with real-world challenges \cite{fang2025social}, where distress may be inevitable but ongoing access to professional or non-professional support is rarely available \cite{stressinAmerica}.

This highlights a key gap between interventions that teach self-care skills through calming experiences versus the practical challenges of day-to-day stress in socially situated contexts. Research in psychology \cite{tulving1973encoding} and applications across domains like nursing and healthcare \cite{rutherford2012learning, drews2013simulation} have supported that learning techniques in simulated, realistic contexts improves skill transfer and recall in applying skills during real-world contexts. The contexts in which we navigate mental health distress similarly involve various constraints of social acceptability and environmental limitation that immersive interventions rarely capture \cite{fang2025social, ross1984socially}. Growing HCI and design literature have begun to propose the importance of situating mental health learning in direct, real-world stress contexts -- thus making the transfer of these skills to the real-world less difficult and nebulous \cite{nunes2015self, nunes2018understanding, wagener2025self}. This work has primarily been theoretical through conceptual frameworks of designing future self-care technology \cite{thieme-hlistic, nunes2015self, wagener2025self}, or qualitative in discovering people's needs and design preferences for contextual practice to train real-world coping \cite{fang2025social}. Yet, empirical and experimental basis for how practicing in a \textit{social simulation} -- the simulation of real-world conditions in which a skill, behavior, or activity is typically performed \cite{fang2025social, dieckmann2007deepening}-- actually affects people's ability to learn, apply, and transfer well-being skills remains unknown. 

As a result, our work builds upon this gap and recent prior work to empirically discover how situating practice in more realistic, contextual environments influences users’ ability to apply and sustain mental health skills. Specifically, our study compares the effects of simulating a real-world stressor environment for practicing mental health skills, compared to conventional isolated or decontextualized environments. In this paper, we answer:

\begin{enumerate}
\item[\textbf{RQ}] \textbf{How does practicing mental health skills in a virtual social simulation affect the application and transfer of skills to the real-world?}
\end{enumerate}

We present findings from a two-week, in-the-wild, between-subject study with 43 participants who use an AR application that teaches calming techniques, applied to the common fear and scenario of public speaking. Our results show that participants who practiced mental health techniques using a `social simulation' intervention -- an AR system that approximates the conditions under which a situated behavior is performed -- showed attenuated rise in physiological stress responses to a mock in-person speaking task as well as greater propensity towards using self-care techniques in their daily lives outside the intervention when presented opportunities to do so, compared to participants who practiced these skills in an intervention sans social simulation. However, participants did not show between-condition differences in self-report stress nor use of the training intervention itself. By presenting this work, we contribute empirical results that situating mental health skill practice in realistic virtual contexts can support aspects of physiological stress regulation and everyday skill transfer, and discuss both opportunities and limitations for the future design of virtual self-care and immersive simulation systems.

%% file: Sections/02-relatedwork.tex
\section{Related Work}
\label{2}

\subsection{Embodied and Immersive Technology for Mental Health}
\label{2.1}

Extended reality (XR) is the extension or replacement of physical reality into a digital environment, and is a promising modality for delivering engaging and controlled forms of mental health support. Immersive technologies have been effective for general well-being such as achieving greater sensory awareness \cite{seabrook2020understanding, bruggeman2018hiatus}, as well as more clinical outcomes such as treating conditions like anxiety and post-traumatic stress  \cite{botella2017recent, reger2016randomized, oprics2012virtual, meyerbroker2010virtual, freeman2008studying}. Relevant to our study, XR systems have been shown to be effective at teaching new mental health skills and coping mechanisms. Immersive environments aid knowledge gain and application of skills by helping users focus and replacing their potentially distracting real-world environment, taking advantage of different sensory inputs, and/or integrating feedback mechanisms based on physiological or behavioral responses \cite{feinberg2022zenvr,prpa2018attending,tan2023mindful, chung2016mindful, wagener2025self}. Approaches to teaching these skills have ranged widely, from gaming to promote learning in an engaging and joyful manner \cite{miner2024stairway, de2023porting, fleming2017serious}, to leveraging physicality and responsive environments \cite{prpa2017pulse, chandrasiri2020virtual, peng2023asmvr} (e.g., allowing users to "physically" push negative thoughts \cite{grieger2021trash}), to more straightforward educational delivery \cite{feinberg2022zenvr}. 

Notably, XR has been effective in exposure therapy-like interventions for repeatedly conditioning people to virtual forms of real-world stressors. For example, XR exposure has been applied across fear-provoking scenarios such as driving \cite{wald2000efficacy}, specific phobias \cite{wechsler2019inferiority, garcia2002virtual}, and social stress \cite{powers2008virtual}. Public speaking is a common application of XR exposure therapy due to its prevalence as a social phobia and its disruptive impact on education, career, and everyday life \cite{lim2023meta, harris2002brief, poeschl2017virtual}. Generally speaking, XR exposure has been found to be roughly as effective as in-vivo exposure therapy \cite{botella2017recent,wechsler2019inferiority, gonccalves2012efficacy}. While these systems primarily aim to reduce stress responses through clinical conditioning, we focus on teaching integration of coping strategies in situ; qualitative evidence has shown that people often face gaps in their current technological and non-technological means to learn self-care practices for enactment in real-world scenarios, and the immersive environment of XR may help fill this gap by providing safe, virtual practical spaces to generate realistic emotional responses from users \cite{fang2025social}.

\subsection{Social Simulation in HCI}
\label{2.2}
Social simulation in HCI has been leveraged frequently in recent years through technology like chatbots or VR/AR in order to understand interpersonal interaction and the psychological and behavioral characteristics of people, such as their self-conception, self-image, or interactions with others \cite{fang2025social, shaikh2024rehearsal, schlagowski2024social, freeman2021body}. With the rapid expansion of immersive text or visual environments across training, therapy, education, and social skills development, researchers have begun to explore how both physical environments and social likeness can be simulated in embodied technology; the integration of large language models (LLMs) have complemented this approach by now offering the ability to generate realistic (and even therapeutic) human dialogue and interaction \cite{lai2023supporting, xu2024mental, shaikh2024rehearsal, yang2024social}. 


Social simulation for mental health has been of particular interest in embodied environments due to the ability to approximate real-world social contexts while maintaining experimental control. Moreover, simulation in the mental health and medical domain has been shown to evoke accurate real-world behaviors and support new skill gain for transferring to real-life contexts; for example, people's behaviors in in-vivo exposure therapy is akin to when presented with virtual reality exposure \cite{powers2008virtual, botella2017recent}, skillsets and decision-making in virtual reality accurately predicts behaviors in real life \cite{wismer2022laboratory, de2018virtual, lloyd2009equivalence}, and community outcomes can be accurately reflected in virtual simulations \cite{moussaid2016crowd}. Drawing on social presence theory, which emphasizes the sense of “being with" another in mediated environments \cite{biocca1997cyborg}, a number of works have shown how embodied environments foster richer interactions than traditional 2D communication platforms \cite{oh2018systematic}; combined with experiential learning theory, we can then see how simulation in embodied environments is a highly useful tool for situated practice alongside abstract knowledge \cite{kolb2014experiential} and an especially well-suited method for studying social behaviors relevant to health and well-being. This continues to be the case even when embodied social simulations are stylized in appearance or behavior \cite{sanchez2005presence, fang2025social, mori2012uncanny}. 

In sum, prior HCI research demonstrates that embodied environments serve as effective hosts for social simulation. Following the assumption across domains such as healthcare, education, and occupational training \cite{bracq2019virtual, bennett2017simulation} that realistic social contexts are critical for practicing occupational skills, the mental health and well-being space may similarly benefit from embodied social simulation and interaction to help users transfer skills to everyday contexts.

\subsection{HCI for Self-Care and Coping Skills}
\label{2.3}
While HCI has made extensive contributions to digital mental health, much of this work has historically focused on professional care (e.g., digital therapy \cite{karyotaki2017efficacy, mohr2017intellicare, schroeder2018pocket}) or peer support \cite{yang2024makes, sharma2018mental, andalibi2016understanding}. Relatively less focus has been on the importance of \textit{self} in mental health care until somewhat recently \cite{nunes2015self, wagener2025self}. Self-care is understood to encompass activities such as self-soothing, emotion regulation, and other practices of self-maintenance \cite{godfrey2011care, levin1983self, riegel2021self}; self-care is a fundamental pillar for maintaining well-being and a necessary condition for success of other types of care \cite{riegel2021self, godfrey2011care, levin1983self}. . 

Recent years have seen growth in the recognition of mental health not only as a clinical concern but as part of everyday life involving self-driven routine practices and self-regulation \cite{nunes2015self, nunes2018understanding, lim2019facilitating}. For example, research has drawn on foundations of self-care in healthcare and psychology, such as through psychological mechanisms like behavior change models and self-determination theory, to design interventions to motivate healthier practices \cite{10.1145/3563657.3596050, 10.1145/3631700.3665241, kim2024my} and bring awareness to one's own well-being \cite{nunes2015self, lee2020toward, yu2018biofeedback, sanches2010mind}. Work has promoted critical design frameworks, future research agendas, and systems contributions for HCI for effective and practical self-care \cite{slovak2023designing,liu2025regulation,yu2018biofeedback, sanches2010mind, fang2025social}. 

Together, this work reflect growing interest in technological intervention for self-care and mental health as a practice of relatively mundane and routine actions that are self-reliant. Nunes et al.'s review of self-care technologies in HCI explicitly surfaces this notion in articulating that the goal of self-care is to be part of practical and everyday actions rather than management of conditions in isolated or solely clinical approaches \cite{nunes2015self, nunes2018understanding}. Similarly, Thieme et al. called for greater approaches to self-oriented, holistic well-being that contrast conventional HCI's focus on treatment or pathology-centered perspectives \cite{thieme-hlistic}, and recent design work by Wagener et al. also proposed a model for self-care emphasizing the importance of repetitive and realistic simulation in line with a technology's ultimate goal \cite{wagener2025self}. Furthermore, qualitative work by Fang et al. found that realistic social simulation in XR could fill a gap in practical, controlled training of mental health practices \cite{fang2025social}. However, despite a growing interest for practical, everyday, and self-directed approaches to mental health along with the use of XR for skill-building, it remains unclear from an experimental point-of-view how implementing more realistic stressor environments for practice  affects intervention use, subsequent stress responses, and everyday enactment of learned skills. 

%% file: Sections/03-hypotheses.tex
\section{Hypotheses}
\label{3}
To explore the influence of  practicing mental health skills in a social simulation environment, we formed four hypotheses investigating (1) \textbf{use} of a social simulation training intervention, (2) \textbf{changes in stress} towards real-world stressors, (3) \textbf{effectiveness} of applying learned skills, and (4) propensity for \textbf{transferring} learned skills to the real-world.

First, a growing body of HCI research suggests that the effectiveness of training and care tools depends heavily on how relevant and meaningful they feel to users \cite{rapp2023exploring, slovak2023designing, jardine2024between}. Interventions that remain too abstract or detached from everyday life risk being perceived as less engaging and less useful \cite{fang2025social}, reducing adherence. Since social simulations mirror the types of stressors people actually encounter in real life, users may have greater motivation to practice skills. We hypothesize that:

\begin{description}
\item [Hypothesis 1: ] \textit{People who practice self-care skills in a virtual social simulation will use the training system more frequently, compared to those who practice in a non-social simulation.}
\end{description}

Second, we evaluate how using social simulation influences participants' stress levels. State, rather than trait, anxiety is defined as “dependent upon both the person (trait anxiety) and the stressful situation” \cite{endler2001state} and is often the measure for fear triggered by specific situations such as public speaking \cite{poeschl2017virtual}. Thus, we hypothesize on how social simulation affects state stress measures. Given that studies generally do not show significant results in a short-term period immediately after a simulation intervention, we hypothesize no differences in state stress after a training intervention using social simulation versus no social simulation \cite{brasil2021stress}. However, practicing with more realistic stressor environments may lead to increased confidence for the user in encountering future real-world stress, and prior work shows that realistic practice results in better preparation and skill acquisition. We hypothesize that participants will report lower \textit{anticipated} stress over time, compared to their baseline stress when starting the study \cite{tanner2012validity}.

\begin{description}
\item [Hypothesis 2: ] \textit{People who practice self-care skills in a virtual social simulation will (a) not show any significant differences in their state stress measures day-to-day, and (b) report significantly lower anticipated stress towards public speaking compared to initial stress levels, compared to those who practice in a non-social simulation.}

\end{description}

Next, we investigate more distal effects such as effectiveness when applying skills and propensity for transferring skills to the real-world. Prior work has found that realistic simulation (both in digital and non-digital ways) is valuable for both immediate and longer-term learning outcomes for applying skills in the real-world \cite{nestel2011simulation, lateef2010simulation}. We build on this to measure whether practice in a social simulation leads users to apply these skills more effectively in real life situations.  We hypothesize that:

\begin{description}
\item [Hypothesis 3: ] \textit{People who practice self-care skills in a virtual social simulation will display lower levels of stress when enacting the mental health skills in real life, compared to those who practice in a non-social simulation.}
\end{description}

Research in psychology and learning sciences has consistently highlighted the importance of situated practice for promoting the transfer of skills from training environments to real-world contexts \cite{tulving1973encoding, rutherford2012learning, drews2013simulation,wrisberg1991effect, michalski2019using}. When learners practice skills in contexts that approximate the physical, social, or emotional constraints of real-life situations, they are more likely to encode relevant cues and adapt strategies effectively when faced with similar challenges outside of the training environment. Given hypothesized effectiveness of social simulation in building mental health skills, we reason that participants who train in a social simulation are also more likely to continue training practice, to their own volition and outside study environments. We hypothesize that:

\begin{description}
\item [Hypothesis 4: ] 
\textit{People who practice self-care skills in a virtual social simulation will show higher propensity for
both \textbf{(a)} voluntarily continuing practice using the training application in their everyday lives and \textbf{(b)} applying skills in their real-world lives, compared to those who practice in a non-social simulation.}
\end{description}

%% file: Sections/04-methods.tex
\section{Methods}
\label{4}
We conducted a 14-day, 43-participant experiment study. In this section, we describe our methods:
\begin{enumerate}
    \item First, we detail key design principles and development of two versions of an AR application: \textit{control} and \textit{treatment}, which participants are randomly assigned to. \textbf{(Section \ref{4.1}).}
    \item Next, we review the experiment design. To assess our research questions, we required consistent user engagement as well as in an in-the-wild manner to assess H4(a). Users thus used the intervention for (a) a required period of 7-days, followed by (b) an optional "in-the-wild" period of 7-days, and finally (c) engaged in an interview and mock speaking task after the 14-days with the application. During the participants' 14-day period with the system, we collected day-to-day insights on their self-reported stress levels, use of the AR system, and transfer of the included mental health skills in the real-world. \textbf{(Section \ref{4.2}).}
    \item Finally, we overview the measures collected and our analysis methods. \textbf{(Section \ref{4.3})}.

\end{enumerate}

\subsection{System Design}
\label{4.1}

\begin{figure*}
 \centering
    \includegraphics[width=\linewidth]{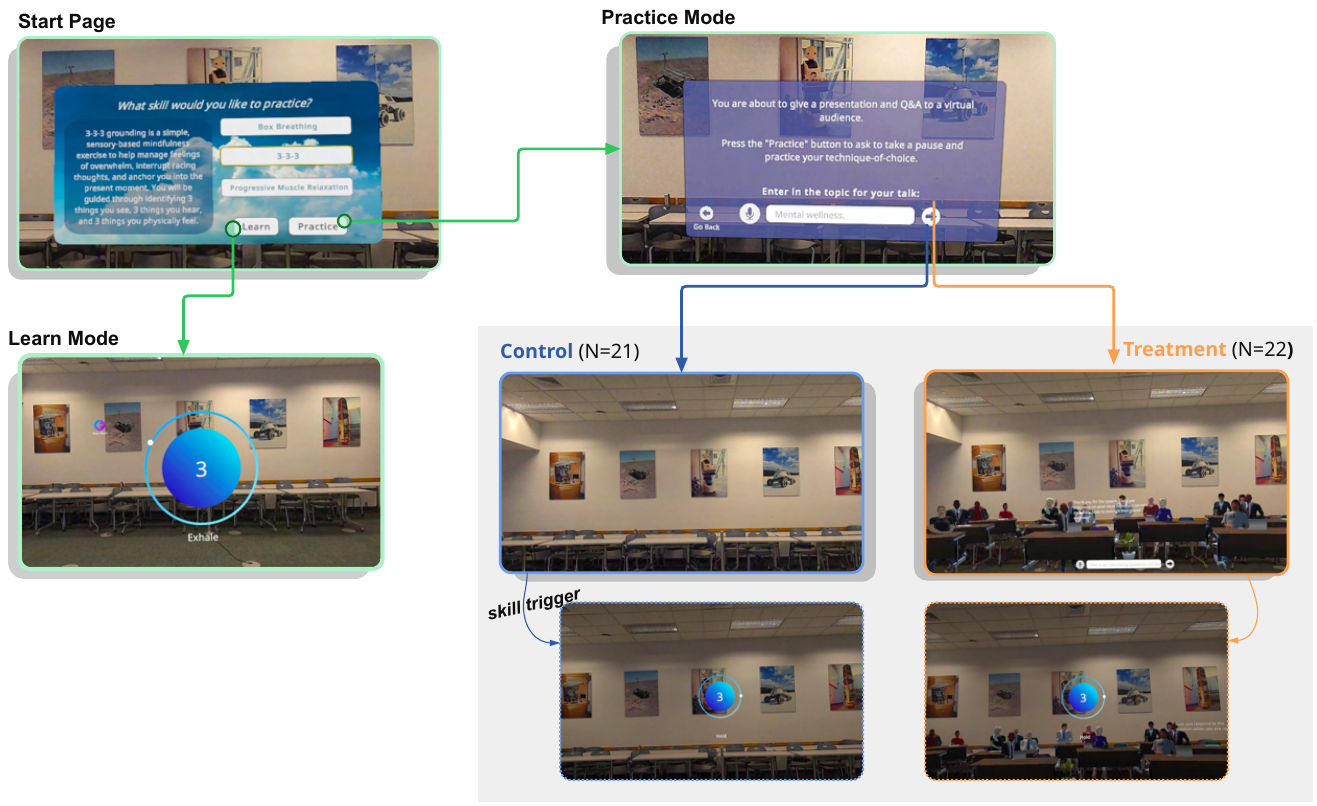}
    \caption{User flow through the control and treatment versions of the AR system.}
    \label{fig:user-flow}
\end{figure*}
To examine the influences of simulating a stressor environment, we built an AR application for the Meta Quest 3. Although distress triggers vary from person-to-person, we designed the application to focus on just one type of scenario to control across participants: \textbf{public speaking}. 

Public speaking is the most common fear \cite{dwyer2012public} and especially relevant to our recruited population (see Section \ref{4.2}); for example, 64\% of college students experience fear of public-speaking \cite{marinho2017fear}. Public speaking is also well established regarding exposure through embodied technology, as it can evoke similar levels of fear and arousal from people as in-person engagements \cite{la2025exploring, chanwimalueang2016modelling, slater1999public}. As a result, we designed the application to act as a practice tool for people to tackle mental health distress while delivering a public talk and Q\&A session for an audience.

To evaluate our hypotheses, we developed an AR application. Below, we describe the full system followed by the distinguishing factors between two versions of the system for our experiment: \textit{control} and \textit{treatment}.

\subsubsection{Overview of System}
\label{4.1.1}
We begin by describing the user flow for the application (Figure \ref{fig:user-flow}).

Upon startup of the application, users are presented with a selection for one of three mental health skills to practice: (1) Box Breathing, (2) 3-3-3, or (3) Progressive Muscle Relaxation. Upon hover, users see a brief description of each of these techniques as well as its intended outcomes. For example, hovering over Box Breathing shows: "\textit{Box Breathing is a simple and effective breathing exercise that promotes relaxation and calms the nervous system. You will be visually guided with an expanding and contracting orb through a controlled, rhythmic pattern.}". All methods (i.e. box breathing, 3-3-3, progressive muscle relaxation) that our team chose to be included in the application have been found to be effective at reducing stress, anxiety, and/or panic, and are easy for beginners to learn \cite{tan2023mindful, carlson1993efficacy, gan2022effects,Nunes-Harwitt_2018}. After the user selects one of the three methods, they may select one of two modes: \texttt{Learn} and \texttt{Practice}. 

\textbf{\texttt{Learn Mode. }} \texttt{Learn Mode} allows the user to familiarize themselves with a selected method without any stressor stimuli. \texttt{Learn Mode} superimposes visual and auditory guidance for practicing the technique, at the center of the attention of the user. For \texttt{Learn Mode}, Box Breathing consists of an expanding and contracting orb that guides the user through a countdown timer for inhaling, holding the breath, exhaling, and holding at the bottom of the breath for four seconds each; 3-3-3 contains text and audio guidance on identifying objects, sounds, and feelings in the user's physical world; and progressive muscle relaxation guides the user through text and audio for relaxing different parts of the body from feet to the head while the user simultaneously follows an inhale-exhale breath pattern through a soundbite guiding their breath through audio. \texttt{Learn Mode} intentionally emulates the design of current self-care applications on market, such as meditation apps.

\textbf{\texttt{Practice Mode. }}  In \texttt{Practice Mode}, users are prompted through both text and audio to give a talk out loud. Since the application is in augmented rather than virtual reality, users can also refer to their own notes and presentation slides if present. In the \textit{treatment} condition, the system begins a visual and auditory simulation of a classroom/meeting room environment filled with audience members. Avatars are seated in the space, accompanied by soft background noises (e.g., shuffling, typing). After the talk, users proceed to a Q\&A session in which virtual audience avatars take turns asking questions, and participants can reply verbally through a speech module (Figure \ref{fig:user-flow}). Throughout \texttt{Practice Mode}, all participants can press a button to bring up their selected mental health skill (box breathing, 3-3-3, progressive muscle relaxation) in an abbreviated, superimposed version if they wish to practice during their talk.

\textbf{Control vs. Treatment Versions. } We create two versions of the above system. What we will refer to as the \textit{control} version of our system is the same as the \textit{treatment} (full) version of the system, but \textbf{with the social simulation elements (virtual audience, augmented classroom objects) removed.} In \texttt{Practice Mode}, control participants were given the same instructions and interactions for conducting a speech and Q\&A but were instructed to `\textit{imagine}' an audience; \textit{treatment} includes the full system features as described above including the virtual audience. Both versions share the same Learn Mode, the speaking/Q\&A structure of Practice Mode, instruction delivery modalities, and the ability to deploy self-soothing skills mid-task, but differed in the presence of the simulated public speaker environment. Since the system was in AR, participants could use the application in any physical environment; however, we encouraged all participants to use the application in an open plan space as much as possible, such as classrooms on campus. 
\begin{figure}[h]
\includegraphics[width=\linewidth]{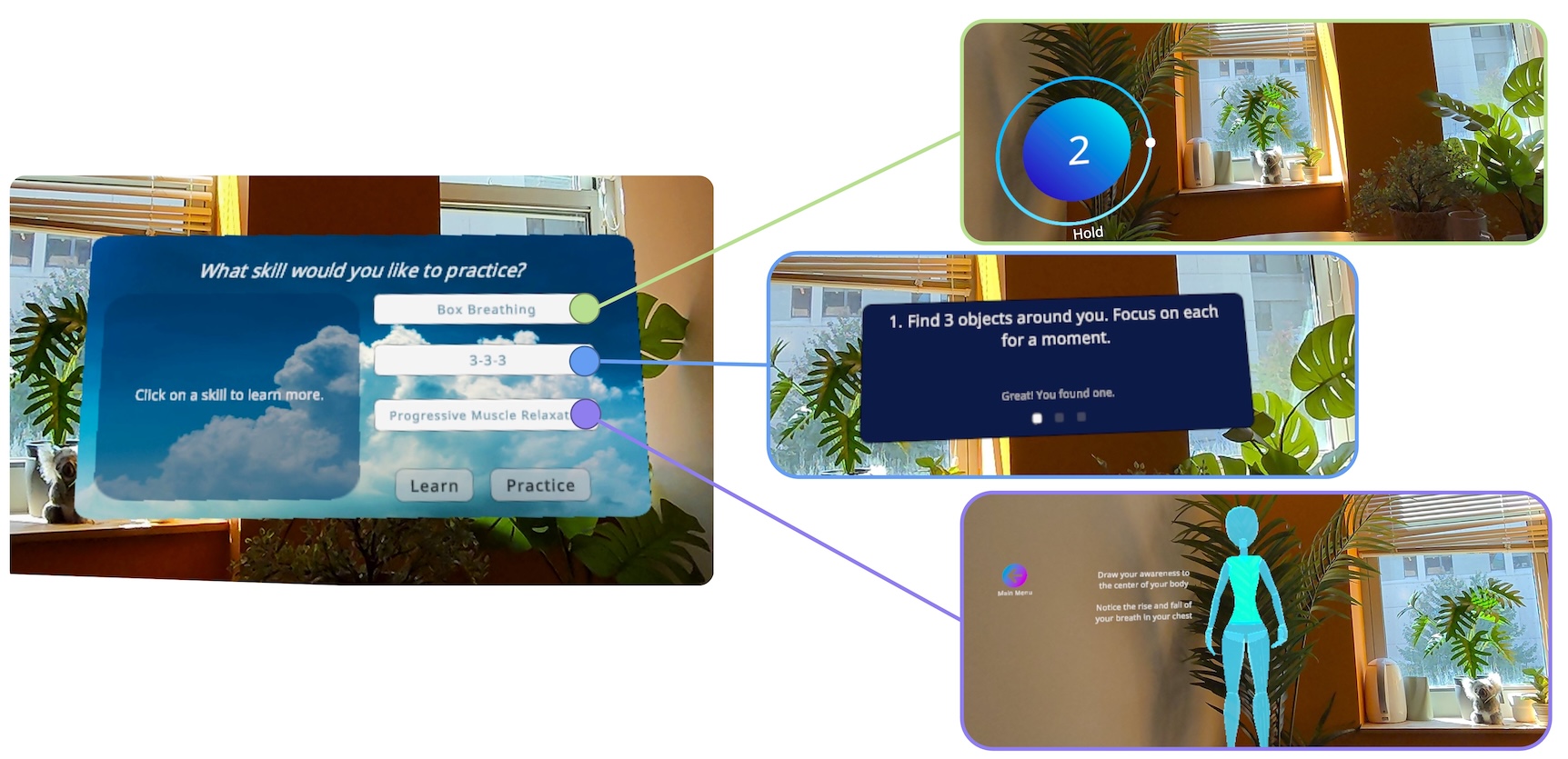}
\caption{Screenshots of the three mental health skills present in our system: Box Breathing (top), 3-3-3 grounding (middle), and Progressive Muscle Relaxation (bottom).}
\label{fig:learn-options}
\end{figure}
We also note that we define \textit{social simulation} as the emulation of the social conditions under which a behavior is performed, which consists of two components: \textit{(1) social interaction} (e.g., engagement with embodied avatars) and \textit{(2) material and environmental context} (e.g., spatial layout, classroom objects, and ambient audio). While definitions of social simulation can vary, we draw most closely from nursing simulations and exposure therapy \cite{drews2013simulation, bouchard2017virtual}, which emphasize that simulating stressful real-world scenarios does not depend only on social actors, but also on the tools, objects, and physical context that may also evoke stress. As such, we designed social simulation in our work as an integrated design configuration, which approximates the social conditions under which a behavior is ordinarily performed. Accordingly, our treatment condition includes the interactive avatars in a classroom-like environment (e.g., desks) and dialogue in order to simulate the social setting, rather than avatars standing silently in unnatural isolation. Teasing apart influence of these individual factors is a potential future work, but out of scope for this study and inconsistent with prior literature on what social simulation and stressor environments are encompassing of. Since the experimental manipulation adds/removes the social simulation altogether, the effects that we investigate should be interpreted as the \textit{effect of the integrated social simulation system} rather than the effects of individual design factors (see Section \ref{6.4}). 

\subsubsection{Design Principles and Implementation}

Below, we overview implementation details guiding design of both control and treatment systems.

\textbf{Modality. } For building a system that could validly examine how users differ in practicing with more realistic, virtually simulated environment, we built an AR application using the Unity game engine and deployed on a Meta Quest 3 headset. We decided to build the environment in augmented reality due to previous work that has found it to meet users' qualitative needs in simulating socially-related stress as users can remain situated in physical surroundings where stress might arise \cite{fang2025social}.

\textbf{Environment. } For \textit{treatment}, we present participants with a simulated audience. All avatars were customized using UMA 2. Additionally, we looped sound of soft classroom and hallway noises to emulate a typical speaking engagement environment. 

\textbf{Interaction. } Dialogue for the avatars was generated using LLMs (GPT), given its ability to generate realistic, responsive, and context-specific replies to the user that more closely emulate real-world interaction. For the treatment condition, we used a prompting pattern (see Appendix A) for avatars to switch turns asking the participant questions on their talk. To ensure natural exchanges, the first and third authors iteratively refined prompting patterns, eliciting diverse tones and question types including those that were less deferential. For example, our initial pilot tests included issues of avatars being too pleasant and unrealistic to the range of tones typically present in a public speaking engagement; we included explicit prompts to rectify these issues (see appendix). Voices were generated through OpenAI’s AudioAI text-to-speech. Participants interacted verbally using simple microphone input, which we then transcribed using OpenAI’s Whisper. Avatars were assigned subtle behaviors like posture shifting.

\textbf{Mental Health Skill Development. } Across both versions, participants could select from three techniques: Box Breathing, 3-3-3 grounding, and Progressive Muscle Relaxation. These methods were chosen as beginner-friendly, evidence-based strategies for reducing distress \cite{tan2023mindful, carlson1993efficacy, gan2022effects,Nunes-Harwitt_2018}. Their design drew on prior literature \cite{lukic2021physiological,tan2023mindful}. Participants in both control and treatment could access the chosen skill at any point during Practice Mode, ensuring strategies were available during stressful moments. This mirrored real-world coping, allowing us to study how users incorporated techniques under stress -- either with a simulation of a social simulation (\textit{treatment}) or an imaginative environment (\textit{control}).

\subsection{Experiment Design}
\label{4.2}

We conducted a 43-person between-subject user study. Participants were randomly assigned to the \textit{control} or \textit{treatment} systems described above. Below, we describe participant recruitment and criteria, followed by each of the four stages of the study.

\subsubsection{Participants}
To ensure safety of our Meta Quest 3 equipment given the large number of participants and in-the-wild nature of the study, we recruited out of our academic institution in an urban area of the Mid-Atlantic region of the United States. All participants had to sign up via their institutional email address, and so are either undergraduate or graduate students, staff, or faculty at our research team's institution. We recruited participants using physical flyers, social media, institutional mailing lists, and snowball sampling. 

\begin{table*}[ht]
\centering
\setlength{\tabcolsep}{4pt}
\begin{tabularx}{\textwidth}{c l Y l c c c l Y l c}
\toprule

\multicolumn{5}{c}{\textbf{Treatment Group}} & \multicolumn{6}{c}{\textbf{Control Group}} \\
\cmidrule(r){1-5} \cmidrule(r){7-11}
\textbf{ID} & \textbf{Gender} & \textbf{Race} & \textbf{Age} & \textbf{Initial SUDS} & & 
\textbf{ID} & \textbf{Gender} & \textbf{Race} & \textbf{Age} & \textbf{Initial SUDS} \\
\midrule

\rowcolor{lightgray}1  & Cis Woman & Asian           & 25--29 & 7 & & 3  & Cis Woman & Asian  & 18--24 & 5 \\
2  & Cis Man   & White           & 18--24 & 6 & & 4  & Cis Woman & Asian  & 18--24 & 4 \\
\rowcolor{lightgray}5  & Cis Man   & Hispanic        & 18--24 & 5 & & 6  & Cis Man   & Asian  & 18--24 & 4 \\
7  & Cis Woman & Asian           & 25--29 & 7 & & 9  & Cis Woman & Asian  & 18--24 & 6 \\
\rowcolor{lightgray}8  & Cis Man   & Asian           & 18--24 & 4 & & 11 & Cis Woman & Asian  & 25--29 & 5 \\
10 & Cis Man   & Asian           & 18--24 & 3 & & 12 & Cis Woman & Black  & 18--24 & 5 \\
\rowcolor{lightgray}17 & Cis Man   & White           & 18--24 & 5 & & 13 & Cis Man   & Black  & 18--24 & 3 \\
18 & Cis Woman   & White  & 18--24 & 3 & & 14 & Cis Man & Hispanic  & 25--29 & 8 \\
\rowcolor{lightgray}19 & Cis Man   & Asian           & 18--24 & 4 & & 15 & Cis Woman & Asian  & 18--24 & 5 \\
20 & Cis Man   & Asian           & 25--29 & 3 & & 16 & Cis Man   & Asian  & 18--24 & 6 \\
\rowcolor{lightgray}22 & Cis Woman & Asian           & 18--24 & 5 & & 21 & Cis Woman & Asian  & 18--24 & 7 \\
26 & Cis Woman & Black           & 18--24 & 7 & & 23 & Cis Woman & Asian  & 18--24 & 5 \\
\rowcolor{lightgray}28 & Cis Woman & Asian           & 18--24 & 8 & & 24 & Cis Woman & Asian  & 25--29 & 3 \\
30 & Cis Woman & Asian           & 25--29 & 2 & & 25 & Cis Man   & Asian  & 25--29 & 3 \\
\rowcolor{lightgray}32 & Cis Woman & Asian           & 25--29 & 2 & & 27 & Cis Woman & White  & 25--29 & 5 \\
33 & Cis Man   & Asian           & 25--29 & 2 & & 29 & Cis Woman & Asian  & 18--24 & 4 \\
\rowcolor{lightgray}34 & Cis Woman & Asian           & 18--24 & 5 & & 31 & Cis Man   & Black  & 25--29 & 5 \\
36 & Cis Man   & Asian           & 25--29 & 7 & & 35 & Cis Woman & Asian  & 25--29 & 3 \\
\rowcolor{lightgray}37 & Cis Man   & Black           & 30--34 & 5 & & 38 & Nonbinary & White  & 18--24 & 2 \\
39 & Cis Man   & Asian           & 35--39 & 7 & & 41 & Cis Woman & Asian  & 18--24 & 7 \\
\rowcolor{lightgray}40 & Cis Man   & Hispanic        & 18--24 & 5 & & 43 & Cis Man   & Asian  & 18--24 & 3 \\
42 & Cis Woman & Asian           & 18--24 & 3 & &     &           &        &        &   \\
\bottomrule
\end{tabularx}
\caption{Participant demographics by condition. Initial SUDS is on the self-report scale of 1 to 10.}
\label{tab:participants}
\end{table*}

Given that we are not focused on self-care for chronic conditions, we recruited a non-clinical population. We define ‘clinical’ and ‘non-clinical’ populations based on mental health literature by recruiting participants who (1) do not have diagnosis of anxiety or stress disorder and (2) are not engaged in any professional treatment to minimize confounding variables with our study's intervention. Upon filling out our interest form, participants are asked to self-score themselves on the Subjective Units of Distress (SUDs) ("On a scale of, how stressful or anxious do you anticipate yourself to be when engaging in public speaking?"). SUDS is an established rating system in psychology \cite{tanner2012validity,kiyimba2020clinical} and was chosen as the most appropriate scale for our use case from a senior professor of psychiatry, who is also the director of a notable center for psychotherapy and has extensive experience as a practicing psychiatrist. We recruited participants who reported having at least mild public speaking anxiety (a score of 2 or higher) on the 10-point scale; we did not recruit participants who report themselves to have extremely severe public speaking anxiety (9 or 10 points) on their self-report, for participants' psychological safety reasons. We also recruited individuals who confirm that they have at least one upcoming speaking engagement in the next two months. This approach allowed us to better understand practice and use of self-soothing techniques for participants who the public speaking application is relevant and generates some amount of real-world stress.  

In the end, we recruited 44 participants with 1 person who dropped out shortly after start of the study (due to unrelated personal circumstance). Participants were compensated \$50 USD through gift card or Zelle. This study was approved by the appropriate Institutional Review Board (IRB). All studies were conducted from June through August 2025.

\subsubsection{Stage 1: Onboarding (Day 0)} 
During a 30-minute in-person onboarding session, participants were asked to fill out a consent form and provide self-report ratings for their familiarity with VR/AR and various self-soothing techniques (“Rate your familiarity with (virtual and augmented reality technology / box or deep breathing techniques / 3-3-3 or grounding techniques / muscle relaxation or body scan techniques)”) on a Likert scale of 1 (Not at all familiar) to 5 (Extremely familiar) \cite{vagias2006likert}. Participants were given a Meta Quest 3 headset, which had assigned (control vs treatment) system according to randomized assignment. Finally, our research team members gave the same 5-10 minute tutorial session that familiarized all participants with the headset controls. 

\subsubsection{Stage 2: Required Use Period (Days 1 through 7)}
Participants were instructed to use the system every day for the next seven days for 10-minutes each day. We did not record the microphone or input content of the participants for privacy reasons, but did record local logs of button presses during their session so we could verify and analyze their daily use. Each day, the participants were instructed to fill out a survey, consisting of the same questions for before and after the participant's application use for the day:

\begin{enumerate}
\item Have you used any self-soothing skills learned from the AR application in your real life (outside of the headset) in the past 24-hours?
\item \textit{(Before/After using the application)} What is your stress level at the current moment?
\item \textit{(Before/After using the application)} What is your anticipated stress level, if you had to conduct public speaking in real life?
\end{enumerate}

The full survey with answer choices is shown in Appendix.

\subsubsection{Stage 3: Optional Use Period (Days 8 through 14)}
For the second week of the study, in order for us to understand participants' free use of the application and incorporation of the application into their daily lives in a non-study context, participants kept the Quest 3 and were told they no longer were obligated to use the system for 10 minutes per day. While participants did not have required intervention use, we still recorded logs of their use if they did use the application, and survey responses.

\subsubsection{Stage 4: Assessment and Semi-Structured Interview (Day 15)}
After two weeks, participants returned to the research site for a mock public speaking task in order for us to measure differences in physiological responses between participants when doing tasks in real life. We also conducted semi-structured interviews to gain qualitative insight.

We drew inspiration from prior literature \cite{douglas_physical_2022} and asked participants to engage in a public speaking task to our research team. All Stage 4 tasks were conducted by the first two authors. Upon first entry to our room on campus, we asked the participants to wear a FitBit Versa for heart-rate measurement. We then let the participant settle into a chair while we recorded participants' starting (resting) baseline BPM. We then asked the participants to imagine that are being interviewed in a room of executives and to speak for about five minutes to answer the question: "\textit{What are your best and worst qualities?}". When necessary (for example, if participants finished answering the question quickly), we necessarily asked participants follow-up questions to extend their heart-rate recording, such as "\textit{Can you elaborate?}", "\textit{How about as a team member?}". We recorded their heart rate throughout this activity, therefore collecting heart rate data before they began answering the question, during the course of their answer, and after they finished speaking. We then asked participants to conduct a second task of conducting a mental health skill taught in the AR system (box breathing, 3-3-3, progressive muscle relaxation) for two minutes. Similarly, we recorded their heart rate before, during, and after their completion of the self-soothing task.

After the two tasks, we conducted 30-minute semi-structured interviews with all participants. The interviews were guided by a list of questions but allowed to deviate depending on topics participants may introduce. Questions included probing on participants experiences of stress over the last few weeks, methods used to tackle these situations, and transfer of skills (e.g., “\textit{What did you do, if anything, to deal with the feelings of stress or anxiety in that situation?}”). We asked about experiences with the application and their experiences navigating practice sessions (e.g., "\textit{How did you decide, if at all, to pause during your speaking or Q\&A session?}"). We voice-recorded all discussions and analyzed transcripts using thematic analysis.

\subsection{Analysis}
\label{4.3}
We first summarize and review the measures and outcomes we gathered from the user study for our analysis.

\begin{enumerate}
\item In Stage 1, we gathered participants' demographics, self-reported familiarity with VR/AR technology and various mental health skills, and starting SUDS score for public speaking. 

\item For Stages 2 and 3, we gathered their daily survey information and logs of use, including duration and frequency of application use.

\item Finally, in Stage 4 we collected participants' heart-rate data during a mock in-person speaking task and self-soothing task, as well as qualitative interviews.
\end{enumerate}

For all measures relevant to our hypotheses (Section \ref{3}), we ran statistical tests and regressions to understand how treatment and other covariates affected outcomes. We also qualitatively analyzed semi-structured interviews of participants through an iterative analytic cycle consisted of two team members (1) recording and transcribing interviews, (2) inductively coding transcripts, (3) amalgamating codes, (4) discussing codes among the entire research team, (5) highlighting themes relevant to each research question, and (6) writing and revising memos \cite{braun2012thematic}.

%% file: Sections/05-results.tex
\section{Results}
To evaluate our hypotheses, we measured the effect of treatment and control assignments for usage statistics (Section 5.1), self-report measures of stress (Section 5.2), effectiveness of applying skills (Section 5.3), and transfer of learned skills to participants' real lives (Section 5.4). We assessed baseline balance between randomized groups on all collected demographics and prior familiarity measures; we found no significant imbalances between treatment and control on any baseline variable (all $p>0.05$).

\subsection{H1: Usage}
\label{results:h1}
\begin{table*}[ht]
\centering
\begin{tabular}{lccc}
\toprule
\textbf{Variable} & \shortstack{\textbf{(a) Total sessions (log)} \\ $\beta$ (SE)} & \shortstack{\textbf{(b) Practice sessions} \\ $\beta$ (SE)} & \shortstack{\textbf{(c) Length of use per session} \\ $\beta$ (SE)} \\

\midrule
(Intercept) & 1.629** (0.462) & -4.940 (4.769) & 8.980*** (1.362) \\
Treatment & -0.512 (0.720) & 7.490 (7.428) & 5.997** (2.093) \\
Initial SUDS & 0.107 (0.081) & 1.983* (0.834) & 0.413 (0.229) \\
Familiarity VR/AR & 0.009 (0.090) & 0.173 (0.930) & 0.465 (0.278) \\
Familiarity breathing & -0.225 (0.146) & -1.664 (1.506) & -0.610 (0.423) \\
Familiarity muscle relax. & 0.028 (0.147) & -0.870 (1.512) & 0.328 (0.409) \\
Familiarity grounding & 0.451* (0.164) & 8.227*** (1.691) & -0.113 (0.384) \\
\midrule
Treatment × SUDS & 0.011 (0.104) & -1.137 (1.077) & -0.190 (0.313) \\
Treatment × Fam. VR/AR & 0.070 (0.160) & -0.000 (1.647) & -1.015* (0.447) \\
Treatment × Fam. breathing & 0.462* (0.218) & 3.215 (2.253) & 1.003 (0.694) \\
Treatment × Fam. muscle & -0.261 (0.223) & -1.025 (2.296) & -0.982 (0.751) \\
Treatment × Fam. grounding & -0.435 (0.265) & -7.958** (2.733) & -0.444 (0.673) \\
\midrule
\textbf{Multiple $R^2$} & 0.404 & 0.601 & 0.110 \\
\textbf{Adjusted $R^2$} & 0.382 & 0.455 & 0.077 \\

\bottomrule
\footnotesize{*** $p<0.001$, ** $p<0.01$, * $p<0.05$}

\end{tabular}
\caption{Regression models for: (a) log-transformed total application uses, (b) raw count of practice sessions (subset of total uses), and (c) length of use in minutes.}
\label{table:all_usage_models}
\end{table*}

\begin{description}
\item [Hypothesis 1] \textit{People who practice self-care skills using social simulation will use the system more frequently.}
\end{description}

\textbf{We did not find evidence to support H1.} 
We calculated the number of sessions that participants completed across their 14-day period. We define a session as starting when the participant launches the application and completes at least one Learn or Practice session. If the application enters sleep mode (either due to inactivity or because the participant manually puts the headset to sleep), subsequent use is deemed as a new session. 

Upon comparison to the log model residuals, the residuals of the raw count of sessions already appeared to be more normally distributed, but its model explained a significantly smaller fraction of variance (adjusted $R^2$=0.07) compared to modeling the log-transformed outcome. Given that log-transforming the outcome improved model fit, although residuals were slightly less symmetric, we report results using the log-transformed outcome as the primary model. 

As seen in Table \ref{table:all_usage_models}, we found that the effect of treatment on total number of sessions over the course of the study did not show significance ($\beta=-0.51$, $p=0.48$). For the length (minutes) of use per session by participants, though, we found that sessions for treatment participants on average were approximately six minutes longer than control participants' sessions ($\beta$=5.99, $p=0.004$). Overall, we did not find support for H1, as results do not show significant evidence that participants actually used the application more frequently.

However, participants' previous familiarity with the technology and/or techniques were strong predictors of their usage behaviors. Familiarity with grounding techniques was positively associated with the number of sessions ($\beta=0.45$, $p=0.01$) as was the treatment by breathing familiarity interaction ($\beta=0.46$, $p=0.043$). Treatment effects are roughly 58\% higher for participants who had one unit higher familiarity with breathing techniques at the start of the study. This booster effect may be due to these users being already at higher propensity to use such systems given existing personal interest. Conversely, for session length, the treatment effect was smaller for participants with higher VR/AR familiarity. Intuitively, this may simply be a result of those less familiar with the technology needing to spend more time on interface navigation.

Upon a deeper look into participants' specific use statistics for Learn and Practice modes in the application, we found that results for the number of \textit{Practice} sessions echoes these previous results. Only interaction effects between treatment and previous familiarity with mental health techniques showing significance while the main effect of treatment showing no significance ($\beta$=7.49, $p=0.32$). However, higher initial SUDS scores had a small positive effect ($\beta$=1.98, $p=0.024$); for a control participant, each 1-point increase in SUDS at baseline is related to an increase of ~2 additional practice sessions, but treatment does not significantly change this effect. 

\subsubsection{Qualitative Findings}
In terms of learning and applying mental health skills from technological tools, participants generally found the application to be more useful for learning skills than other self-care technologies like meditation apps or video tutorials; however, both control and treatment participants described using \texttt{Learn Mode} only for the first days of the study given that techniques were easy to learn. Treatment participants consistently emphasized that \texttt{Practice Mode} was the more useful feature due to environmental immersion ("\textit{The best way to evoke a certain emotion in a situation is to put the person in that situation}" (P5), "\textit{70\% I would use the app [in the future], because it's helped me a lot to present and reflect...it helped me a lot to get used to the atmosphere}" (P7), "\textit{Before, I used to practice by just saying out loud to myself, but I feel like that's different. Simulating that experience presenting on the spot, I feel like that was good to practice with}" (P26)). Some control participants expressed positivity towards the tool in contrast to traditional methods of learning self-care techniques, such as P16 who commented, "\textit{I often find a lot of situations where I learned something from online, from a video, but it's hard for me to process those steps even if the content creator thinks they are already slow enough, detailed enough. It's not just enough...but VR put me in that position}"; however, some also felt that VR/AR was similarly effective to video or other modalities, while being less accessible given inconvenience of the headset. As a result, some participants expressed that the VR/AR environment needed additional contextual features to feel worthwhile, e.g., P4 (control) said, "\textit{I'm not a good imaginative person, I can't imagine myself in this situation. But with AR, it could be helpful and able to put me in that situation}" and P25 (control) similarly expressed, "\textit{I do wish it had more applicability ... I feel like, if it could collaborate with me, if it could give me, "okay, this is it in a public speaking pattern". I would use that application then}".

\subsection{H2: Self-Report Stress Measures}
\label{results:h2}
\begin{description}
\item [Hypothesis 2] \textit{People who practice self-care skills using social simulation will \textbf{(a)} not show any significant
differences in their state stress measures day-to-day, but \textbf{(b)} report significantly lower anticipated stress towards
public speaking compared to initial stress levels, compared to those who practice in a non-social simulation.}
\end{description}

We fit linear mixed-effects models to examine whether condition was associated with participants’ self-reported stress. \textbf{We did not detect a difference in state-stress change (H2(a))} and \textbf{did not find support for lowered anticipated stress (H2(b))} between control and treatment participants over time. Results of the model are shown in Table \ref{table:stress_models_combined}, where we calculate state and anticipated stress change as the outcome -- i.e. participants' stress levels post-application use, minus their level pre-application use -- as well as treatment and day interaction for trajectory over time. The model included fixed effects for treatment group (intervention vs. control), mean-centered day of the study for which the self-report was filled out, mean-centered baseline stress, and their interactions. A random intercept for participant ID accounted for repeated measures across sessions. We chose not to include familiarity variables, gender, race, or age to maintain parsimony and preliminary models indicated that these demographic variables did not improve the model nor significantly alter treatment estimates. 

\begin{table*}[ht]
\centering
\begin{tabular}{lcc}
\toprule
\textbf{Variable} & \shortstack{\textbf{(a) State stress} \\ $\beta$ (SE)} & \shortstack{\textbf{(b) Anticipated stress} \\ $\beta$ (SE)} \\
\midrule
(Intercept) & -0.589*** (0.115) & -1.058*** (0.211) \\
Treatment & 0.170 (0.164) & 0.202 (0.302) \\
Day (centered) & 0.037 (0.025) & 0.080 (0.043) \\
Starting SUDS (centered) & -0.038 (0.074) & -0.214 (0.136) \\
Treatment × Day (centered) & -0.038 (0.035) & -0.074 (0.061) \\
Treatment × Starting SUDS & 0.080 (0.098) & 0.062 (0.181) \\
\bottomrule
\footnotesize{*** $p<0.001$, ** $p<0.01$, * $p<0.05$}
\end{tabular}
\caption{Linear mixed-effects models predicting session-by-session stress differences: (a) state stress differences and (b) anticipated stress differences toward public speaking. Both models include fixed effects for treatment, mean-centered day, baseline stress, and their interactions, as well as a random intercept for participant ID.}
\label{table:stress_models_combined}
\end{table*}

\subsubsection{Qualitative Findings} To understand these findings, we turned to qualitative results from the participant interviews. \textbf{Treatment participants' indeed mentioned that engaging with the system in some cases either did not change -- or even raised -- their level of stress (both state and anticipated)}. Participants stated that their practice sessions surfaced their hesitancy or lack of preparation for their talk; these resulted in greater awareness of weaknesses in their preparation and self-consciousness, although this was not necessarily brought up as a negative outcome. For example, P10 (treatment) reflected, “\textit{it has been a long time since I spoke in public at all. When I first used the app, I was like, whoa. I’m kind of rushing}”. Similarly, P8 noted, “\textit{Due to the questions that got deeper and deeper, that would make me a little bit nervous. In terms of learning, it’s a positive experience. In terms of self confidence wise, it was a little bit worse}”. Participants also described the simulation as physiologically realistic, sparking the same nerves they might experience in real-world public speaking: “\textit{I actually did feel my heart rate rise. And I was like, what’s going on? This is fake. But in the moment, this is simulating what’s actually going to be like in real life}” (P17, treatment).

Interestingly, we found that participants expressed that the system gave them tools and "\textit{resources to call on}" (P36, treatment) to deal with stress when it inevitably arises in the future, even if it did not \textit{reduce} their anticipated stress. For example, P37 (treatment) stated, “\textit{I have some good tools in my toolkit now to apply if I ever get too wound up}" while P27 (control) stated, “\textit{I guess I’m more confident now because I didn’t have any good tools before -- I just accepted my fate}". Similarly, P10 (treatment) explained, “\textit{I don’t know if I’m going to be less stressed about public speaking, but even if I do feel stressed, I’ll be able to manage that well with the techniques}". Echoing, P39 (treatment) stated: "\textit{Because of the practice I got, I feel I will be slightly more confident. I still don't feel that I'm good at public speaking. But I would at least be able to give it without panicking about it}". 

Although we did not find significant between-condition differences in anticipated stress measures, one key distinction that emerged was how treatment participants described a reframing of stress specifically tied to \textit{social} and \textit{judgment}-related aspects of public speaking, which we did not see mentioned by any control participants. This may reflect treatment participants' exposure to a virtual audience, which gave them more opportunities to confront the contextual and social dimensions. For instance, P30 (treatment) explained that they felt less afraid to answer questions, stating, “\textit{In one of the tries [in the application] I just said ‘I don’t know the answer to this’ just to see what happens...we moved on to the next question...and I was like, guess what? I can do this in real world too. When I’m asked something, I can always say, ‘listen, I don’t have the answer to this today'}”. Similarly, P26 (treatment) reflected on their anxiety towards any silence in their talk and said, “\textit{I feel like using this app has made me more aware of the fact that I can pause, like during a poster presentation...the app kind of reminded me that I can take a pause and take a breath sometimes, if I get nervous when I’m speaking}".

\subsection{H3: Physiological Stress}
\label{results:h3}
\begin{table*}[ht]
\centering
\begin{tabular}{lcccc}
\toprule
\textbf{Variable} 
  & \shortstack{\textbf{(a) Heart Rate over}\\ \textbf{Full Speaking Task} \\ $\beta$ (SE)} 
  & \shortstack{\textbf{(b) Pre-Peak} \\ \textbf{Heart Rate} \\ $\beta$ (SE)} 
  & \shortstack{\textbf{(c) Post-Peak} \\ \textbf{Heart Rate} \\ $\beta$ (SE)} \\
\midrule
(Intercept)                        & \textbf{52.31* (17.31)}   & 35.59 (20.31)          & \textbf{81.55** (2.787)}  \\
Treatment                          & -0.037 (3.451)            & 2.703 (3.80)           & -1.53 (5.236)             \\
Time (sec)                         & \textbf{0.053*** (0.002)} & \textbf{0.131*** (0.003)} & \textbf{-0.061*** (0.003)} \\
Time (sec)$^2$                     & \textbf{-0.0003*** (2.2e-5)} & ---                 & ---                       \\
Initial BPM                        & 0.537 (0.219)             & \textbf{0.623* (0.258)}   & 0.355 (0.352)             \\
Treatment $\times$ Time            & \textbf{-0.034*** (0.002)} & \textbf{-0.033*** (0.004)} & \textbf{0.022*** (0.004)} \\
Treatment $\times$ Time$^2$        & \textbf{0.0001* (3e-05)}  & ---                    & ---                       \\
\midrule
\bottomrule
\multicolumn{4}{l}{\footnotesize{*** $p<0.001$, ** $p<0.01$, * $p<0.05$}} \\
\end{tabular}
\caption{Mixed-effects models predicting heart rate for (a) trajectories across the full speaking task, (b) pre-peak heart rate analysis during speaking, and (c) post-peak heart rate recovery during the speaking task. Note that model (a) includes quadratic time terms, while models (b) and (c) are linear models for each participant's pre- and post-peak (respectively) heart rate to help interpretation. All models include a random effect for participants and predictors: treatment (binary), time in seconds, initial heart rate at start of task, and treatment $\times$ time interaction.}
\label{tab:hr_models}
\end{table*}

\begin{description}
\item [Hypothesis 3] \textit{People who practice self-care skills using social simulation will display lower levels of stress when enacting the mental health skills in real life.}
\end{description}

We analyzed how participants perform when doing a mock in-person speaking task for five-minutes to our research team (as described in Section \ref{4.2}). We recorded their heart rate at baseline before starting the speaking task, during the speaking task, after completing the task, and during a self-soothing activity immediately following. Overall, we found that \textbf{Hypothesis 3 was partially supported} by our results as treatment participants exhibited \textbf{slower modeled rise in heart-rate during the stress task} leading up to peak heart rate and \textbf{faster decline of heart-rate during the self-soothing task}; however, peak heart rate did not differ significantly.

\begin{figure}
\includegraphics[width=\linewidth]{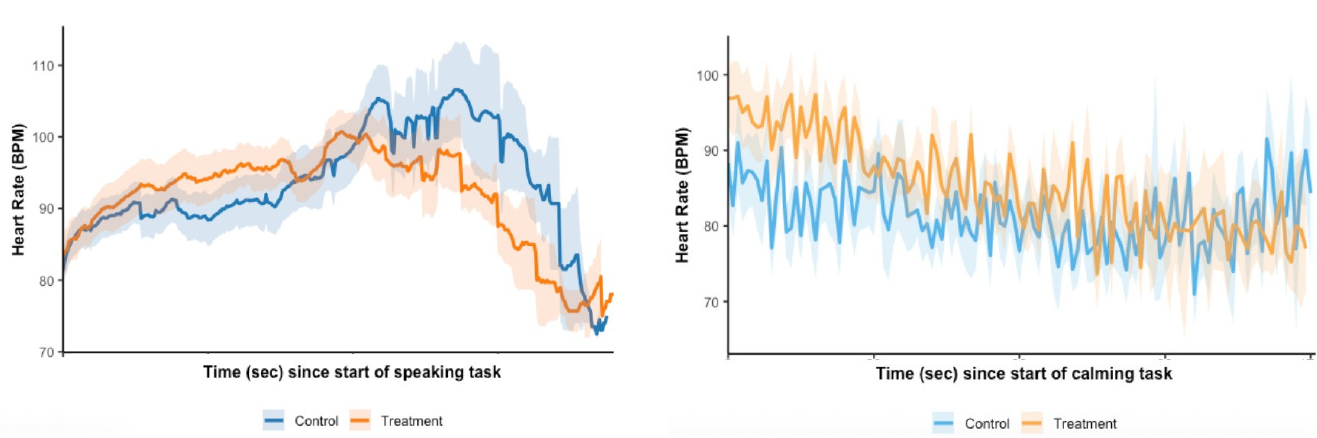}
\caption{Average heart rate trajectories over time for participants in the speaking task for five minutes (left) and during a calming technique task for two minutes (right). Lines represent group means for Control (blue and light blue) and Treatment (orange and light orange), with shaded ribbons indicating ±1 standard error. The plots show the difference in heart-rate rise between treatment and control groups across time.}
\label{fig:heartrate-charts}
\end{figure}

Using a linear mixed-effects model with participant-level random intercepts and accounting for heart rate over the course of the task, we examined the effects of treatment and its interaction with time (sec), controlling for initial resting heart rate. We also ran models using demographic variables, previous familiarities, and total minutes spent using the application over the course of the study as covariates, but these resulted in minimal effect and no significant results for main or interaction effects with treatment. The full speaking task model included quadratic time terms to capture non-linear heart-rate trajectory (see Table \ref{tab:hr_models}). We found treatment and time interaction was significant ($\beta$=-0.034 BPM (beats-per-minute)/sec, $p<0.001$), indicating differences in the modeled trajectories between conditions. 

As a result, we analyzed participants' heart rate more directly leading up to the peak of their heart rate during the task (pre-peak). On average, the peak time of heart rate occurred at 143 seconds for control participants and 141 seconds for treatment participants, with a t-test showing no significant difference in time to peak ($p=0.918$). Using a linear mixed-effects model, results indicated a significant main effect of time and a significant treatment and time interaction, suggesting that treatment indeed slowed the rate of heart rate increase. Initial heart rate was also a significant covariate, indicating a higher `resting' rate was associated with higher heart-rate levels during pre-peak period. Overall, participants who practiced using the treatment system experienced a more moderated slope of heart rate rise both throughout the speaking task and leading up to participants' highest heart rate point, even though peak heart rate itself did not differ between groups.

Additionally, both mediation analyses for whether the length of use (minutes) mediated the treatment effect for (1) average heart rate and (2) slope of heart rate increase showed no significance. We also ran exploratory analyses using HRV measures (RMSSD and SDNN); model results are shown in Appendix for reference. However, we note that HRV is typically done with millisecond-level recordings, so this finding likely reflects some limited resolution of our heart-rate recordings. 

\begin{table}[h]
\centering
\begin{tabular}{lc}
\toprule
\textbf{Variable} & \shortstack{\textbf{Heart Rate During Calming Task} \\ $\beta$ (SE)} \\
\midrule
(Intercept)                 & \textbf{99.08* (21.14)}   \\
Treatment                   & 5.490 (3.96)              \\
Time (sec)                  & \textbf{-0.043*** (0.004)} \\
Initial BPM                 & -0.187 (0.268)            \\
Treatment $\times$ Time     & \textbf{-0.020** (0.007)} \\
\midrule
\bottomrule
\multicolumn{2}{l}{\footnotesize{*** $p<0.001$, ** $p<0.01$, * $p<0.05$}} \\
\end{tabular}
\caption{Linear mixed-effects model predicting heart rate recovery during the calming task. Random intercepts for participants are included.}
\label{tab:hr_calming}
\end{table}

During the calming task, all participants’ heart rate tended to decrease over time. However, treatment participants exhibited a steeper modeled decline in heart rate than control participants ($\beta_{\text{treatment}\times\text{time}}$=-0.020 BPM/sec, $p=0.003$), corresponding to an additional modeled decline of approximately 2.4 BPM over the task. Neither initial heart rate nor the main effect of treatment was significant. These results indicate a difference in recovery trajectories, but we did not find evidence of different absolute heart rate..

\subsection{H4: Transferability}
\label{results:h4}
\begin{description}
\item [Hypothesis 4] \textit{People who practice self-care skills using social simulation will show higher propensity for both \textbf{(a)} voluntarily continuing practice using the training application in their everyday lives and \textbf{(b)} applying these skills in their real-world lives, compared to those who practice in a non-social simulation.}
\end{description}

\begin{table*}[h]
\centering
\begin{tabular}{lcc}
\toprule
\textbf{Variable} & \shortstack{\textbf{Number of Reported Uses of} \\ \textbf{Mental Health Skills in Real Life} \\ $\beta$ (SE)} \\
\midrule
(Intercept)                              & 1.475 (0.824)  \\
Treatment                                & -2.554 (1.520) \\
Initial stress SUDS (centered)           & 0.314 (0.242)  \\
Opportunities for use (binary)           & -1.492 (0.918) \\
Treatment $\times$ Initial SUDS          & -0.345 (0.304) \\
Treatment $\times$ Opportunities for use & \textbf{3.341* (1.614)} \\
\midrule
\bottomrule
\multicolumn{2}{l}{\footnotesize{*** $p<0.001$, ** $p<0.01$, * $p<0.05$}}
\end{tabular}
\caption{Negative binomial regression predicting participants’ real-world use of learned mental health skills. Estimates are log counts with standard errors.}
\label{tab:nb_reported_uses}
\end{table*}

\begin{table*}[h]
\centering
\begin{tabular}{lc}
\toprule
\textbf{Variable} & \shortstack{\textbf{Number of Voluntary Application}\\\textbf{Uses During Optional Period (Stage 3)} \\ $\beta$ (SE)} \\
\midrule
(Intercept)                              & 0.100 (1.859) \\
Treatment                                & -0.062 (0.838) \\
Initial SUDS (centered)                  & -0.698 (0.470) \\
Familiarity: Breathing                   & -0.167 (0.439) \\
Familiarity: Grounding                   & -0.464 (0.553) \\
Familiarity: Muscle                      & 0.086 (0.389) \\
Treatment $\times$ Initial SUDS (cent.)  & 0.205 (0.549) \\
\midrule
\bottomrule
\multicolumn{2}{l}{\footnotesize{*** $p<0.001$, ** $p<0.01$, * $p<0.05$}}
\end{tabular}
\caption{Model predicting participants’ application use during the optional second week.}
\label{tab:lmm_voluntary_uses}
\end{table*}

\textbf{H4(a) was not supported}. We did not find between-condition differences in voluntary application use during the optional period (Table \ref{tab:lmm_voluntary_uses}). In other words, treatment participants did not show significant differences in incorporating greater level of practice (through the training intervention) of mental health skills in their own lives more than control.

We found \textbf{conditional support for H4(b)}; treatment participants reported greater real-world technique use than control participants, \textit{when they reported having opportunities to apply the skills}. We ran a negative binomial model (Table \ref{tab:nb_reported_uses}) on whether participants used learned mental health skills in their real life, self-reported every day during the study. Participants could indicate each day whether they used a technique presented in the intervention. We centered numeric predictors and initially fitted generalized linear mixed models for a binary outcome of whether a participant used a technique present in the intervention ("relevant technique") on a given day. Additionally, we coded participant post-study interviews when we asked explicitly (yes/no) whether and when they had a stressful event in the previous two weeks of the study (which we refer to as an “opportunity for use”), regardless and without prompting of whether the participant used a technique during this event; two coders disagreed on only one ambiguous case, resolved through discussion. We first modeled day-by-day use to examine whether condition was associated with technique use on an average day, but did not find a significant effect (likely due to high variability within days and limited statistical power). We complemented this analysis with a participant-level model examining total reported technique use as an aggregate across the 14-day study, to determine treatment's effect on the total count that a participant enacted techniques in their everyday life.
Given the count outcome and preliminary analysis showing overdispersion, we modeled the data using a negative binomial regression. 

Results are shown in Table \ref{tab:nb_reported_uses}. The treatment-by-opportunity interaction was significant, indicating that among participants who reported having opportunities to do so during the 14-day study, the model estimated 2.2 times as many reported uses in treatment participants compared to control (at mean initial SUDS). Overall, participants in the treatment condition did not increase technique use overall but these results suggest they may have greater transfer for people who encounter situations in which the skills could be applied. We therefore find conditional support for H4(b).

\subsubsection{Qualitative Insight}
When it came to the use of the practice application during the optional period of Stage 3 (Hypothesis 4(a)), nearly all participants stated in their interviews that they did not feel motivated to use the application during the optional use period due to feeling having adequate practice from the 7-day initial use. We did not notice any qualitative difference between treatment or control participants; it is plausible that the application may not have been engaging enough for participants to have incorporated practice into their daily lives regardless of its usefulness for skill-building.

For using the learned mental health skills in their real lives (Hypothesis 4(b)), treatment participants' qualitative perspectives echoed quantitative results, reporting substantially higher rates of applying skills in real life. Several participants described adapting techniques in variations or customized to their particular circumstances such as P26 (treatment) who stated, "\textit{I wasn't necessarily using the breathing techniques that were on the app, but I was doing something similar}" while others described increased awareness of breath (P5), direct application during presentations (P30, P17, P7, P40), or more general daily use (P10, P22, P19). Treatment participants applied skills during public speaking events, such as:

\begin{quote}
"\textit{I did use one in real life recently when I was giving a presentation for my internship...I actually took a minute to stop and breathe just because I was being asked a ton of questions all at once and I was struggling to keep up.}" (P30, treatment)\newline

"\textit{I had two public speaking opportunities during this [research study]. There were times I was like, Oh, maybe I can take a second, especially at the start...a big point of the study that I took away was that I did use these techniques almost in real time...I was subconsciously at a poster presentation and I got a little riled up. Let me breathe for a second. So that was cool to see it play out in real time.}" (P17, treatment)\newline

"\textit{I actually did some breathing because I was super stressed. Public speaking is a big deal for me}" (P40, treatment) 

\end{quote}

as well as in non-speaking related engagements:

\begin{quote}
"\textit{Sometimes, if I'm in a conversation and I'm just listening, I'll randomly just start doing [3-3-3]}." (P22, treatment)\newline

"\textit{Just random times, in the morning, when I reached the office, maybe at night. I would just do box breathing. Or if I'm just sitting idle, I do the muscle relaxation. Those wouldn't signify a stressful moment... I was just like, Okay, why not just do the technique more?}" (P10, treatment)\newline

"\textit{For sort of mentally resetting before or after work, [box breathing] has helped a lot}" (P19, treatment)\newline

"\textit{I woke up and saw a rat. The first thing I did was take deep breaths. I wouldn't say it was exactly like in the practice system but it was very close. I took a couple of deep breaths, calmed myself down. It was like a minute of complete silence and breathing. And it really helped.}" (P30, treatment)
\end{quote}

Surprisingly, despite the lack of social simulation in the intervention, many control participants actually noted that \textbf{the skills were closely tied to the scenario of public speaking in the headset}; as a result, it felt difficult to transfer skills beyond the intervention. For example, P6 stated, "\textit{it's like, closely associated to a task for me, you know, like I have to do this before public speaking}" while P9 said, "\textit{I think I kind of separate the headset from the rest of my life. I didn't really incorporate the two together, and just thought of it as a separate thing.}" Similarly, P16 stated, "\textit{I don't connect it to the real life yet}". Other control participants stated that they were not able to transfer the skills to real-world stress situations due to the, counter-intuitively, stress they felt in those moments. For example, P38 stated, "\textit{I feel like this week has been really stressful...it's the time that I need stress relief the most, but then I neglected during these times}".

In terms of barriers to enacting the skills in their real lives, though, both treatment and control participants described challenges of time and social awkwardness in front of audiences. As P12 (control) explained, "\textit{Real life, you're not gonna be given that chance to try to calm yourself down, Somebody's gonna look at you like, what are you doing?}". As a result, our findings echo \cite{fang2025social} in that the social norms were still a barrier for participants in feeling comfortable to apply self-soothing; as we discuss in Section 5.2.1, though, treatment participants expressed that the intervention helped reframe some social dimensions. It is possible that this is unique to the public speaking application, and other scenarios such as interpersonal conflict with close relationships or self-soothing for general stress could allow people to take pauses more liberally. However, we found that several participants found "workarounds" for applying techniques even within the public atmosphere of public speaking, such as shortening the visible time spent on techniques and using skills before/after speaking rather than integrated within. For example, P10 (treatment) described, "\textit{In a real life situation where I am giving a talk, I don't think I would take 10--20 seconds off to do box breathing...But I definitely do think that I'll be more confident in asking people, Hey, can I take a few seconds off...I would be pretty comfortable using it, but for how long I use it, I probably keep it to a minimum}".

%% file: Sections/06-discussion.tex
\section{Discussion}
Our findings provide an evaluation of a social simulation system for self-care skill practice, emulating stressor environment through embodied social actors and environmental context. Together, our results suggest that a more situated environment for practice can support some aspects of enactment of self-soothing skills.

\subsection{Opportunities for Embodied Social Simulation for Self-Care Skills}
Our findings suggest that mimicking stressor environments through social, environmental, and interactive factors can be a meaningful way to enable self-care skill practice. Participants assigned to social simulation showed a slower modeled pre-peak heart-rate rise during speaking and a steeper decline during the subsequent calming task. We did not detect corresponding differences in practice frequency or within-session self-reported stress changes. Additionally, qualitative results showed that the enactment and transfer of skills even extended beyond the particular application of public speaking. As a result, we find that social simulations can benefit the application and generalization of self-care skills. 

Additionally, we found qualitative support that participants who practice skills using social simulations were more likely to reframe how they anticipated approaching stress in the future; interestingly, participants who practiced with the social simulation (in this case, public speaking) did not seem to be limited to the scope of applying skills to the training context, but rather expressed many different generalizations of these skills to other aspects of daily life. For HCI, this highlights a design opportunity: contrary to intuition, \textbf{using an “anchor scenario” not only trains coping in that domain but may also scaffold broader transfer}, and perhaps even more so than the practice of a mental health skill in a generalized but isolated virtual environment. Designers may wish to select application contexts that are common, evocative, and relatable, enabling users to feel the usefulness of skills but also abstract strategies into their broader lives.

One challenge our study surfaces is how to design interventions that deliberately \textit{resist} awe or spectacle. While fantastical XR environments are undoubtedly effective for engagement and can also lead to profound mental health effects \cite{miller2023awedyssey}, our findings and other prior work \cite{fang2025social, nunes2015self, nunes2018understanding} suggest that technology for \textbf{mundane self-care} can help people create tangible change in their everyday for well-being. Designing for the mundane can raise the issue of engagement; although our quantitative findings found that treatment participants used the intervention longer, indeed a few participants in our study suggested various elements of gamification (e.g., P5 who stated they wanted it to be "\textit{more like a video game}", P16 who suggested "\textit{reward incentives}") to help them continue use -- arguably, features that may also run antithetical to the concept of designing for the practical and mundane everyday. One promising design direction for future research may be the incorporation of "reflective" incentives. Rather than points, badges, or streaks, interventions can highlight meaningful progress (e.g., showing users how their stress response has changed over time, prompting short reflections on users' lived challenges) or lightweight check-ins that align with mundane rhythms (during a commute, after a meeting) so that practice feels embedded in daily life. Of course, other options can create varied experiences to avoid user boredom like customization features (e.g., changing audience sizes, social dynamics) or difficulty levels, which still preserve relevance.

\subsection{Social Barriers in Adoption of Self-Care Skills from AR Simulation}

While participants found value in the AR system, they also articulated barriers that complicate the adoption of self-soothing in everyday contexts. A central theme was the role of \textbf{social pressure}. While participants in both control and treatment expressed that they felt comfortable doing the techniques in private moments and executing the techniques to high proficiency, several participants described that taking any amount of pause to enact self-soothing would be perceived as disruptive or socially unacceptable. As we mentioned, several participants thus created ``lightweight'' versions of skills specifically to fit into social norms. As a result, we echo the implications of previous work in the domain of social simulation for self-care \cite{fang2025social} in that the effectiveness of training people for mental health skills is not only a matter of ability or learning gain, but also \textit{broadly shaped by norms of interaction}. Our qualitative results highlight that social stigma and contextual constraints remained reasons many participants hesitated to transfer skills to their real-life events. While offering a social simulation for treatment participants seemed to spark some reframing around social factors of stress (see Section \ref{results:h2}), our intervention and study were not intensive nor lengthy enough for participants to likely become significantly more comfortable with deviating from social norms. 

\subsubsection{Working Within vs. Challenging Social Norms}
There are two directions that can be considered in future embodied self-care tools: working \textit{within} social norms or \textit{challenging} social norms.

One way to work \textit{within} social norms is assuming the reality that one cannot conduct full-scale self-soothing in social settings, and so instead centering micro-practices -- short, subtle actions that can be enacted quickly and without drawing social attention. Participants highlighted the difficulty of long pauses or visible exercises during public speaking or conversation. Designing for less conspicuous practices (e.g., a single deep breath, quick muscle release) can lower the barrier to real-world use by making self-soothing techniques easier to integrate fluidly into everyday moments of stress. Another design possibility for addressing this is using immersive technology for learning and training of skills, but integrating with wearable device or mobile notification that provide discreet prompts for triggering the practice of these skills in real life. For example, haptic pulses from a smartwatch could gently remind users but also guide a single breathing cycle without requiring them to close their eyes, gesture visibly, or interrupt conversation. These cues build a bridge between formal practice in AR and real-world enactment, reminding users of the techniques when they are most needed. 

However, another direction is designing for \textit{challenge} of the social stigma around self-care. For example, one might design features that also focus on incorporating conversational skills for the user to suggest a break for themselves in a difficult moment (e.g., simulating various reactions to a user asking a virtual audience member if they can take a moment). This can be an empowering vision for addressing and reducing stigma for self-care, both for self and in social contexts. A related example of challenging these assumptions around stigmatizing practices of self-care may also be designing interventions for \textit{group} dynamics, intended to bring shared self-care practices \textit{into} social dynamics. Designing for shared experiences such as group breathing exercises at the start of a meeting, classroom, or performance rehearsal can normalize and reframe well-being as communal rather than individual. From these suggestions, we advocate for future self-care technologies meant for training well-being skills to centrally consider in their design \textbf{the spectrum at which they accept versus challenge the social environment and cultural expectations} of enacting well-being practices.


Together, the above social barriers suggest that adoption of self-soothing techniques and the attempted scaffolds for this through technological means requires more than effective skill training; instead, it also depends on navigating social norms and broader cultural beliefs about the relationship between technology and well-being.

\subsection{Productive Forms of Distress in Mental Health Technology}
We did not find differences between conditions in participants’ state or anticipated stress changes. Our findings suggest that the value of contextual simulation is not necessarily in reducing reaction to a particular stressor, but rather as a useful design to help users practice recognizing and responding to distress under conditions that approximate where the skills must eventually be applied.

Historically, self-care work in HCI has been evaluated through the lens of stress reduction as a positive outcome. Applications such as mindfulness tools and relaxation games often measure success by how much they lower users’ immediate anxiety, heart rate, or subjective stress. This framing assumes that comfort is the goal and the metric of progress. However, our work and those similar \cite{fang2025social, benford2012uncomfortable} may instead promote "productive discomfort”; rather than aiming for soothing experiences, systems may deliberately introduce authentic stress to catalyze growth, self-awareness, and skill development \cite{benford2012uncomfortable, halbert2015designing}. As Halbert and Nathan describe in their work, designs can spawn critical reflection and transformative learning from the experience of distress \cite{halbert2015designing}. Although designing uncomfortable experiences is not new  in HCI, the most prevalent applications have been where maximizing comfort is obviously inappropriate and explicitly counterproductive -- addressing deep disparity and colonialist history \cite{das2022collaborative, dosono2020decolonizing}, engaging in debate \cite{halbert2015designing}, and sparking entertainment through physical or psychological discomfort \cite{benford2012uncomfortable}. Our work suggests that \textit{even in the field of mental health and self-care -- a field marked by the particular desire for comfort -- designing with discomfort and distress in mind (and even as a goal) can also be beneficial for well-being}. Even in interventions meant to soothe, discomfort can serve a vital role in offering a safe environment to practice response. Participants in our study explicitly mentioned that they did not necessarily feel \textit{reduced stress} towards public speaking, but rather felt more \textit{capable} of dealing with the inevitable distress that they expected to have. Participants’ interviews illustrated how discomfort led to growth as they learned to normalize pauses, admit when they lacked answers, and confront judgment with less fear. These outcomes represent a shift away from stress reduction and toward stress literacy \cite{varlow2009stress}.



\subsection{Limitations}
\label{6.4}
We acknowledge several limitations in our work. First, we have limited insight into the exact topics and practice content of participants during their 14-day in-the-wild study. For privacy purposes and for participants to feel more comfortable, we did not record audio or content of participants' interactions beyond behavioral logs described. As a result, confounding variables may exist pertaining to the content that participants chose during practice and the mock speaking task in Stage 4. However, we note that participants were randomly assigned to conditions with no a priori reason to believe that treatment participants systematically chose different practice content than control participants. Future experiments may benefit from opt-in content logging or post-session reflections to capture richer practice data. Second, we acknowledge that the definition of social simulation may vary across work. Our treatment condition included both social interaction and environmental context (e.g., classroom objects). We consider both components to be constitutive of realistic social simulation and the stress it triggers, consistent with prior work noted in Section \ref{4.1.1}. While disentangling the variables of social actors and physical environment may be possible in future work, social stress in everyday life arises not from any one of these elements in isolation; as a result, our work can be considered a first step towards understanding more realistic environments versus isolated environments, but not as an isolation of individual design factors. Third, we cannot establish longer term effects past the duration of our study (14-days). Fourth, much of our analyses rely on self-report data (e.g., daily stress surveys, self-reported application of mental health skills), which can be susceptible to biases like social desirability and recall bias. Future studies could mitigate these limitations through triangulation with methods such as passive sensing (with appropriate consent practices). Finally, we note that our participant pool was drawn from a single academic institution in a U.S. urban area. This limits the generalizability of our findings to broader populations such as older adults or those with limited prior exposure to these technologies.

%% file: Sections/07-conclusion.tex

\section{Conclusion}
We investigated whether an integrated AR simulation using a more realistic stressor environment for practice supports mental health skill practice. In a 14-day randomized study with 43 participants, participants who practiced using the social simulation condition (treatment) showed slower modeled pre-peak heart-rate rise during a mock speaking task and a steeper decline during a calming task. We did not detect between-condition differences in practice frequency or self-reported stress changes. However, treatment participants qualitatively described generalizing the use of learned skills even beyond the training context (public speaking). Overall, these results provide empirical support for designing embodied interventions that embed simulated real-world interaction, showing that practicing skills in realistic contexts can enhance both physiological and experiential outcomes. 